\documentclass[a4paper,nofootinbib,preprintnumbers,floatfix,superscriptaddress,pra,twocolumn]{revtex4}
\usepackage{amsmath, amsthm, amssymb}
\usepackage{graphicx}
\usepackage{dcolumn}
\usepackage{bm}
\usepackage{color}
\usepackage{amsmath}
\usepackage{amsfonts}
\usepackage{amssymb}
\usepackage{hyperref}

\newcommand{\eqnref}[1]{(\ref{#1})}
\newcommand{\figref}[1]{Fig.~\ref{#1}}
\newcommand{\secref}[1]{Sec.~\ref{#1}}

\newcommand{\vacr}{|vac\rangle}
\newcommand{\vacl}{\langle vac|}
\newcommand{\bra}[1]{\langle #1|}
\newcommand{\ket}[1]{|#1\rangle}
\newcommand{\braket}[2]{\langle #1 | #2 \rangle}
\newcommand{\ketbra}[3][]{\ket{#2}_{#1}\!\bra{#3}}      
\newcommand{\upa}{\uparrow}                             
\newcommand{\dwa}{\downarrow}                           
\newcommand{\thr}{^{\text{crit}}}						

{\begin{framed}\begin{small}}
{\end{small}\end{framed}}

\DeclareGraphicsExtensions{.pdf,.png,.jpg}

\makeindex

\begin{document}

\title{Feasibility of loophole-free nonlocality tests with a single photon}
\date{\today}

\author{Rafael Chaves}\affiliation{ICFO-Institut de Ci\`encies Fot\`oniques, Mediterranean Technology Park, 08860 Castelldefels (Barcelona), Spain}
\author{Jonatan Bohr Brask}\affiliation{ICFO-Institut de Ci\`encies Fot\`oniques, Mediterranean Technology Park, 08860 Castelldefels (Barcelona), Spain}

\begin{abstract}
Recently much interest has been directed towards designing setups that achieve realistic loss thresholds for decisive tests of local realism, in particular in the optical regime. We analyse the
feasibility of such Bell tests based on a W-state shared between multiple parties, which can be realised for example by a single photon shared between spatial modes. We develop a general error
model to obtain thresholds on the efficiencies required to violate local realism, and also consider two concrete optical measurement schemes.
\end{abstract}

\maketitle

\section{Introduction}\label{sec.Intro}Local realism -- the assumption that physical quantities have well
established values previous to any measurement and that signals
cannot travel faster than the speed of light -- entails
limits on the correlations which space-like separated,
independent observers may obtain. Such restrictions, usually expressed as
Bell inequalities~\cite{Bell1964}, may be surpassed within quantum
theory when the observers share certain entangled quantum states. The correlations
then cannot be explained by any local hidden variable (LHV) model,
and are hence labeled nonlocal correlations.

The experimental evidence for the existence of nonlocal
correlations is compelling yet not completely conclusive. Existing
tests all suffer from one or more loopholes, which open up LHV
explanations of the data unless further assumptions are
introduced. In some cases, the measurements are not sufficiently
fast or far apart to be space-like separated
events~\cite{Bell_Ions}. In other cases, low detection efficiency
is the obstacle. For example, in Bell tests based on photon
pairs~\cite{Aspect1982}, low photodetection efficiencies imply
that in many experimental runs, at least one photon is lost.
Analysing only the coincidence counts, nonlocal correlations may be
observed even when the total dataset can be explained by a LHV
model. An additional fair-sampling assumption is therefore
required to reject local realism in experiments with low detection
efficiency. While this may seem natural, albeit slightly
dissatisfactory, in experiments probing the nature of quantum
mechanics, it is incompatible with device independent applications
for which the violation of a Bell inequality is a necessary
condition to assure e.g.~security of certain tasks~\cite{Acin2007,
Pironio2010,gerhardt2011}. For example, a malicious eavesdropper could take
advantage of detection inefficiency to break a device-independent
key distribution protocol without being detected.

The robustness of Bell test violations to loss and inefficiencies depends on the particular test and setup. To achieve efficiency thresholds compatible with those of laboratory detectors, various approaches have been considered, e.g.~changing the number of settings, outcomes and parties~\cite{Pitowski1989, Massar2002, Brunner2007, Cabello2008, Vertesi2010, Pal2009}, as well as the detection schemes and types of states~\cite{Cabello2007, Brunner2007, Ji2010, Cavalcanti2011, Quintino2011, Sangouard2011}. Interestingly, as first noted by Eberhard~\cite{Eberhard1993}, the state providing better robustness against losses, is not necessarily the most entangled. For the CHSH inequality, corresponding to 2 parties, 2 settings, and 2 outcomes, a critical efficiency of 67\% is achieved by an almost separable state. However, such a state is very susceptible to noise, and hence not terribly practical in experiment.

Recently, asymmetric setups have attracted attention. Asymmetry in
the efficiency of different measurement settings is motivated by
the fact that the measurements which can be efficiently performed
do not necessarily coincide with those yielding a high Bell test
violation. For example, homodyne detection of light can be
extremely efficient, but it seems very difficult to obtain good
violations based solely on homodyning~\cite{Patron2004}. Combining
homodyning and single-photon detection, in
Ref.~\cite{Cavalcanti2011} the authors obtain a threshold of 71\%
for single-photon detection. This is higher than for the Eberhard
scheme, however the required state is not close to separable and
seems feasible to prepare. Asymmetry between parties is motivated
by the natural occurence of entanglement between different
physical systems, such as atoms and photons, for which the
available detection schemes have widely different efficiencies.
Atomic states can be measured with almost unit efficiency.
Nonlocality tests for atom-photon systems have been investigated
in Refs.~\cite{Cabello2007, Brunner2007}, where a critical
single-photon detection efficiency of 45\% was obtained, and very
recently in Ref.~\cite{Sangouard2011}, where 39\% was obtained for
a scheme combining atoms, single-photon and homodyne detection.

In this paper, we consider general asymmetric Bell tests, based on
multipartite W-states \cite{Wstate}. The physical implementation
we have in mind is a single photon shared between multiple
parties, and possibly entangled with an additional atomic system.
However our analysis is applicable also to a wider qubit setting.
The motivation for focusing on nonlocality tests based on W-states
is threefold: the simplicity of preparing single-photon W-states,
the robustness of W-state entanglement against losses, and the
existence of Bell inequality violations which rapidly approaches
the algebraic maximum for an increasing number $N$ of parties.

In the following sections we first discuss optical W-states and examine concrete measurements schemes in \secref{sec.Physical}. We introduce a POVM model capturing a broad range of detection imperfections and loss in \secref{sec.POVM}, and in Secs. \ref{sec.Bell} and \ref{sec.Polytope} we explore the robustness of nonlocality for the W-state by means of specific Bell inequalities and linear programming respectively. We demonstrate, using the Bell inequality proposed in Ref. ~\cite{Heaney2011}, that the local content of the W-state tends exponentially fast to zero for increasing $N$. This, and the fact that the entanglement in the state has a very robust (size-independent) resistance to losses~\cite{Chaves2010}, would seem to suggest that there could be an advantage to increasing $N$ in tests of nonlocality. However, we show that this is not the case for several concrete examples where the robustness actually decreases with $N$, while in other cases which we have examined, only slight improvements with $N$ are found. At the same time the threshold scaling, for the cases with no improvements, is not severe. This is positive, since multipartite nonlocality is important in its own right and useful in several information processing tasks~\cite{CCP}.

\section{State and physical measurements}

\label{sec.Physical}The W-state for $N$ parties, each a 2-dimensional quantum system, can be taken to be
\begin{equation}
\label{eq.W} \ket{W_N} = \frac{1}{\sqrt{N}} \left( \ket{10\ldots0}
+ \cdots + \ket{0\ldots01} \right) ,
\end{equation}
where $\ket{0}$, $\ket{1}$ denote the eigenstates with eigenvalues $1$, $-1$ respectively of the Pauli operator $\hat{\sigma}_z$. In a purely optical implementation, \eqnref{eq.W} can be created
by coherently distributing a single photon among $N$ parties. Heralded single photons can be generated e.g.~via spontaneous parametric down conversion, and distributed by means of linear optics. Measurements of $\hat{\sigma}_z$ then correspond simply to single-photon detection (SPD). However, measurements in the $x$-$y$ plane of the Bloch sphere require transformations such as $\ket{0}+\ket{1}
\rightarrow \ket{0}$ that do not preserve energy, and hence cannot be implemented perfectly with passive, linear optics. Nevertheless, approximate implementations are possible as discussed below.

Another case we will consider is atom-photon entanglement. For single trapped atoms or ions, very high detection efficiency and good control over the measurement bases can be achieved \cite{Bell_Ions}. At the same time, through spontaneous emission, the atom can be entangled with an optical field, e.g.~in a state of the form $\cos(\theta)\ket{e,0} + \sin(\theta)\ket{g,1}$, with
the first mode referring to the atomic state and the second to the number of photons in the field. Distributing the emitted photon over multiple modes, one arrives at a W-state with one atomic
party and $N-1$ photonic parties, that is
\begin{equation}
\label{eq.Wasym} \ket{W_{asym}} = \cos(\theta)\ket{e}\vacr_{N-1}+\sin(\theta) \ket{g}\ket{W}_{N-1},
\end{equation}
where $\vacr$ denotes a state with all modes in $\ket{0}$. Since the coupling $\eta_{c}$ between the atom and the emitted photon is not perfect, the actual state will be an incoherent sum of $\cos(\theta)\ket{e}\vacr_{N-1} +\sqrt{\eta_{c}} \sin(\theta) \ket{g}\ket{W}_{N-1}$ and $\ket{g}\vacr_{N-1}$ with respective probabilities given by $\cos^2(\theta)+\eta_{c}\sin^2(\theta)$ and $(1-\eta_{c})\sin^2(\theta)$.

\subsection{Pauli measurement via homodyning}
\label{subsec:homo}

An approximate implementation of $\hat{\sigma}_x$ can be achieved using the fact that in the 0-1 photon Fock space, a homodyne measurement with phase $\phi$ and sign-binning approximates a measurement of $\cos(\phi)\hat{\sigma}_x+\sin(\phi)\hat{\sigma}_y$, i.e.~in the equatorial plane of the Bloch sphere \cite{Quintino2011}. For example, within the 0-1 photon Fock space, the elements of the projection operator approximating $\hat{\sigma}_x$ are
\begin{align}
\pi^x_{nm} &= \bra{n} \left( \int_{0}^{\infty} dx \ketbra{x}{x} - \int_{-\infty}^{0} dx \ketbra{x}{x} \right) \ket{m} \nonumber \\
& = \int_{0}^{\infty} dx
\psi_{\ket{n}}(x)\psi_{\ket{m}}^*(x) - \int_{-\infty}^{0} dx
\psi_{\ket{n}}(x)\psi_{\ket{m}}^*(x) \nonumber \\
& =\sqrt{\frac{2}{\pi}}\bra{n}\hat{\sigma}_x\ket{m} .
\end{align}
Due to the non-unit factor $\sqrt{2/\pi}$ the detection is not perfect. The probability to produce the correct output given either eigenstate of $\hat{\sigma}_x$ is $\frac{1}{2}(1+\sqrt{2\eta_{hom}/\pi})$, where $\eta_{hom}$ is the homodyne detection efficiency. For unit efficiency this number becomes $\sim 89.9\%$.

\subsection{Pauli measurement via displacement and SPD}
\label{subsec:dis}

In Ref.~\cite{Laghaout2010} an alternative method was proposed for the implementation of measurements of $\hat{\sigma}_x$. In this setup the incoming field is displaced by mixing with a coherent state on a highly transmitting beam splitter and subsequently measured with a single-photon detector. For a real displacement $\alpha$, the probability to observe no clicks at the detector (in the absence of loss), when an eigenstate of $\hat{\sigma}_x$ is incident, is given by
\begin{align}
\label{eq.dispP0}
P(0|\pm) & = \left| \bra{0} D(\alpha) \left(\ket{0}\pm\ket{1}\right)/\sqrt{2} \right|^2 = \frac{1}{2} |\braket{\alpha}{0} \pm \braket{\alpha}{1}|^2 \nonumber \\
& = \frac{1}{2} e^{-\alpha^2} \left( 1 \pm \alpha \right)^2 ,
\end{align}
where $D(\alpha)$ is the displacement operator and $\ket{\alpha}$ denotes a coherent state. Thus, by choosing $\alpha = -1$ one can ensure that the state $\ket{0}_x$ is faithfully detected, in the sense that it will always give a detector click, while the state $\ket{1}_x$ will give a no-click outcome with probability $2/e$. As we will see in the next section this has the same structure as an SPD $\hat{\sigma}_z$-measurement. However, this choice of $\alpha$ suggested in Ref.~\cite{Laghaout2010} is not necessarily optimal. When the single-photon detector has a non-unit efficiency $\eta_{spd}$, \eqnref{eq.dispP0} becomes
\begin{equation}
\label{eq.dispP0loss}
P(0|\pm) = \frac{1}{2} e^{-\eta_{spd}\alpha^2} \left( (1 \pm \eta_{spd}\alpha)^2 + 1 - \eta_{spd} \right) .
\end{equation}
The optimal choice of $\alpha$ will depend on $\eta_{spd}$ as well as on the Bell scenario in which the approximate $\hat{\sigma}_x$-measurement is used.

\section{General error analysis}
\label{sec.POVM}

In any experiment, imperfections are likely to be present in the form of losses, detector inefficiencies, and noise. Depending on the particular setup, these imperfections might have different effects. We imagine Bell experiments in which each party chooses between measurements of the same two dichotomic observables, labeled by $s$ and $s'$. For most cases we will consider, the settings $s$, $s'$ will correspond to implementations of $\sigma_z$ and $\sigma_x$ respectively. A broad range of error behaviour can be modeled within the same framework, describing each measurement by a POVM with elements
\begin{equation}
\label{eq.povm}
\begin{split}
M_\uparrow & = \eta_\upa \ket{\upa}_s\bra{\upa} + (1-\eta_\downarrow) \ket{\dwa}_s\bra{\dwa} , \\
M_\downarrow & = \eta_\dwa \ket{\dwa}_s\bra{\dwa} +
(1-\eta_\upa)\ket{\upa}_s\bra{\upa} ,
\end{split}
\end{equation}
where $\ket{\upa}_s,\ket{\dwa}_s$ are the eigenstates of the observable $s$. For ideal efficiencies $\eta_{\dwa,\upa} = 1$ this POVM implements a projective measurement along the direction defined by $s$. For non-ideal efficiency we capture several types of errors as we now describe in detail.

One common loss-induced error occurs in experiments based on single-photon detection, where imperfect detectors lead to a decrease in the probability for observing one outcome (click) and an increased probability for the complementary outcome (no click). This makes sense e.g.~for an optical implementation of \eqnref{eq.W}, with the $z$-basis corresponding to the Fock basis and $\ket{\upa}_z = \ket{1}$. The vacuum never leads to clicks (in the absence of dark counts) while the single photon is detected with a finite efficiency. We can model this by taking
\begin{equation}
\label{eq.err_spd}
\eta_\dwa = 1 \text{\ and\ } \eta_\upa \in [0,1] .
\end{equation}
Note that this models also describes the approximate implementation of $\sigma_{x}$ via displacement, for $\alpha= \pm 1$, as discussed in Sec.~\ref{subsec:dis}.

Another relevant error consists in each input state being incorrectly identified as its opposite with some small probability. This corresponds for example to a polarisation measurement on a single photon, with a slight misalignment of the experimental measurement basis. As discussed in Sec.~\ref{subsec:homo}, it also occurs when homodyne measurements are used to approximate a measurement of $\sigma_x$. We can model this error by
\begin{equation}
\label{eq.err_sym}
\eta_\dwa = \eta_\upa \in [0,1] .
\end{equation}

Combining the POVMs \eqnref{eq.err_spd} and \eqnref{eq.err_sym}, several interesting scenarios can be described. For example, taking \eqnref{eq.err_spd} with $\eta_\upa = \eta$ for $\hat{\sigma}_z$-measurements and \eqnref{eq.err_sym} with $\eta_\upa = (1+\sqrt{\eta})/2$ for $\hat{\sigma}_x$-measurements models an experiment where the state is subject to an amplitude damping (AD) channel which incoherently replaces $\ket{1}$ by $\ket{0}$ with probability $1-\eta$, and where all parties measure $\hat{\sigma}_z$, $\hat{\sigma}_x$. Such an AD channel could describe e.g.~transmission loss affecting a single-photon W-state. Another example is the dephasing (D) channel, which is relevant for trapped-ion experiments, where dephasing is caused by magnetic-field and laser-intensity fluctuations, or spontaneous emission during Raman transitions \cite{ion_review}, and also for photonic polarization-qubits \cite{optics_review}. For $\hat{\sigma}_z$- and $\hat{\sigma}_x$-measurements, the D channel with phase-flip probability $1-\eta$ is modeled by \eqnref{eq.err_sym} with $\eta_\upa = 1$ and $\eta_\upa = (1+\eta)/2$ respectively.

One can imagine other scenarios for which measurements return a result that does not correspond to any of the binary outcomes, e.g.~no click does not correspond to any polarization direction.
To treat such an additional outcome one can either adopt a binning strategy, grouping it with one of the binary outcomes, or one can consider a setting with more outcomes, for which two-outcome Bell inequalities then no longer apply. Binning can be modelled within the POVM framework. For example \eqnref{eq.err_spd} can be seen as binning of no-click events with the $\dwa$-outcome. We will return to additional outcomes in \ref{sec.Polytope}.

\section{Bell inequalities approach}
\label{sec.Bell}

In this section we analyze the feasibility of the $N$ modes W-state and possibly entangled with an additional atomic system to perform loophole free nonlocality tests using several Bell inequalities.

The atomic party is assumed to be able to perform ideal measurements in any basis. In order to find the best thresholds we optimise over these bases  and also on the coefficient $\theta$
that describes the atom-photon entangled state (\ref{eq.Wasym}). The photonic parties are assumed all to perform the same measurements: single-photon detection with efficiency described by eq.
(\ref{eq.err_spd}) or an approximate implementation of $\cos(\phi)\hat{\sigma}_x+\sin(\phi)\hat{\sigma}_y$. Actually, in all the cases considered, we found numerically that the best thresholds for the efficiencies are obtained for $\phi=0$, that is for $\hat{\sigma}_x$ measurements.

\subsection{Cabello et al. inequality}

To begin with we investigate the inequality proposed for the 3-qubit W-state by Cabello in Ref.~\cite{Cabello2002} and later generalised to more parties in Ref.~\cite{Heaney2011}. Writing $p(o|s)$ for the probability that outcome $o \in \{0,1\}$ is obtained with the measurement setting $s \in \{\hat{\sigma}_x,\hat{\sigma}_z\}$, in our variant the inequality takes the form
\begin{widetext}
\begin{equation}
p(\mbox{all 0}|\mbox{all  $z$}) + p(\mbox{all but one 0}|\mbox{all $z$}) - p(\mbox{$x$'s different, all $z$'s 0}|\mbox{two $x$, all else $z$}) - p(\mbox{all equal}|\mbox{all $x$}) \leq 0 ,
\end{equation}
or more formally $B \leq 0$ with
\begin{align}
\label{eq.ineq}
B = & p(0,\ldots,0|z,\ldots,z)  + \sum_\nu p(\nu(1,0,\ldots,0)|z,\ldots,z) - \sum_{\nu'} p(\nu'(1,0,\ldots,0)|\nu'(x,x,z\ldots,z)) \\
& - p(0,\ldots,0|x,\ldots,x) - p(1,\ldots,1|x,\ldots,x) , \nonumber
\end{align}
\end{widetext}
where $\nu$ runs over all cyclic permutations of length $N$ and $\nu'$ runs over all permutations giving distinct images of $(1,2,0,\ldots,0)$. We note that the first term on the left-hand
side was not present in previous works \cite{Cabello2002,Heaney2011,Laghaout2010}. However, it is easy to check that the inequality is still valid, and since the term is always positive it can only increase any violation of the inequality. The violation of \eqnref{eq.ineq} attained by an $N$-party W-state approaches 1 exponentially fast with $N$, specifically
\begin{equation}
\label{eq.heaneybound}
B(\ket{W}) = 1-\frac{N}{2^{N-1}}.
\end{equation}

\begin{figure}
\includegraphics[width=.48\textwidth]{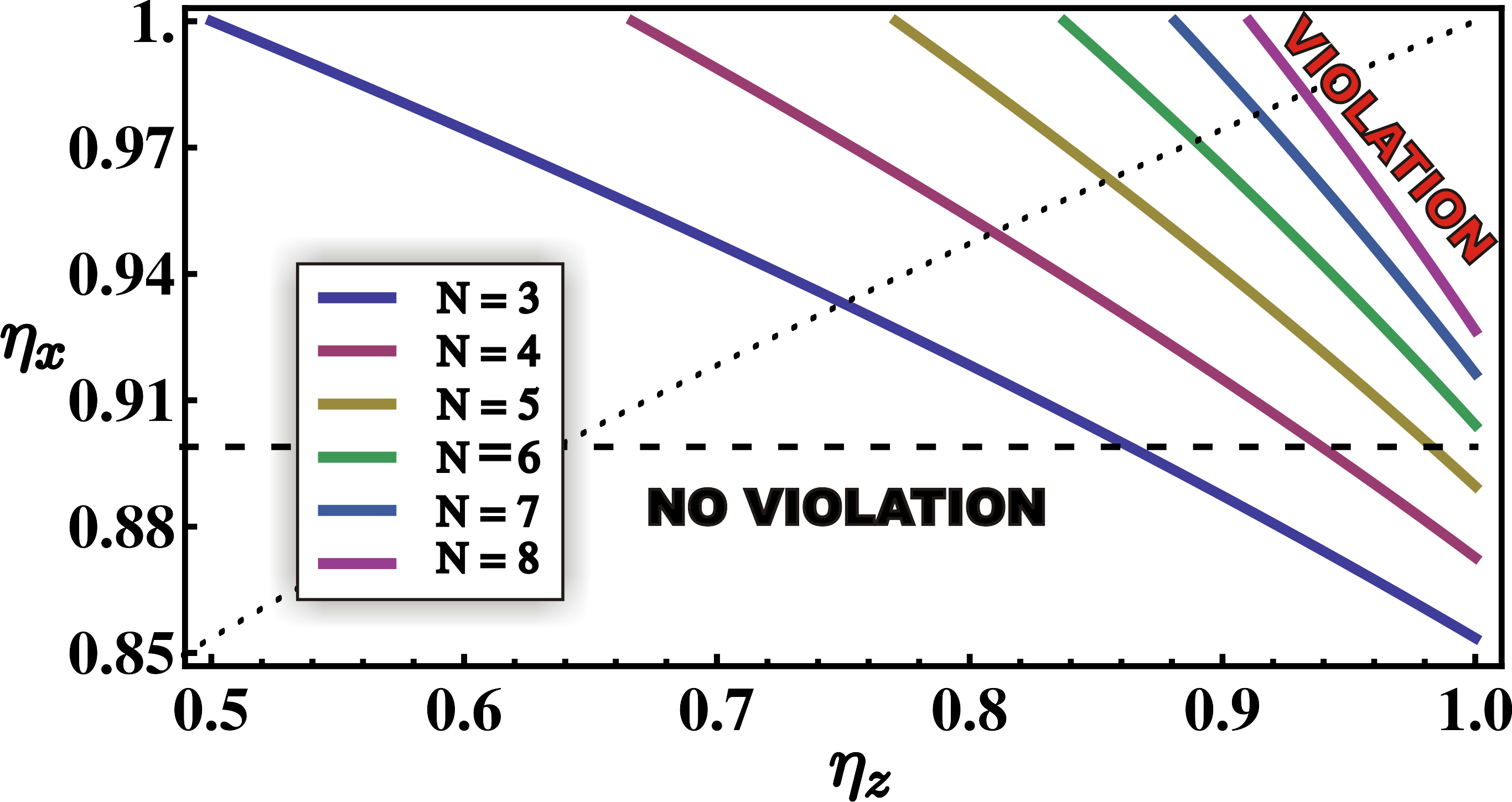}
\caption{Regions of violation of the Cabello inequality \eqnref{eq.ineq} for $N=3,\ldots,8$ (bottom to top), when the error behaviour of $\hat{\sigma}_z$- and $\hat{\sigma}_x$-measurements is given respectively by \eqnref{eq.err_spd} and \eqnref{eq.err_sym}. The inequality is violated above the solid lines. Intersections of the dotted and dashed with the solid lines give respectively the critical survival probability for the AD channel and the critical $\hat{\sigma}_z$-efficiency when $\hat{\sigma}_x$ is approximated via homodyning.}\label{fig.povm_spd_hom}
\end{figure}

In Ref.~\cite{Cabello2002,Heaney2011} no analysis of the robustness of the violation was presented. From the growing violation and the robustness of the W-state entanglement against
losses one might speculate that there could be experimental advantage to increasing the number of parties. As an illustration consider the $N$-party W-state undergoing amplitude damping in each of its qubits, thus becoming
\begin{equation}
\label{eq.wampdamp}
\Phi_{amp}(\ket{W}\bra{W}) = \eta \ket{W}\bra{W} + (1-\eta) \vacr\vacl ,
\end{equation}
where $\vacr$ denotes the state with all modes in $\ket{0}$, and $1-\eta$ is the damping rate (or equivalently the loss probability). To assess the effect of the damping, we need to compute the value of $B$ for the state $\vacr$, and we find
\begin{equation}
\label{eq.vacbound}
B(\vacr) = 1 - \frac{1}{2}\left( \begin{array}{c} N \\ 2 \end{array} \right) - \frac{1}{2^{N-1}} .
\end{equation}
From \eqnref{eq.heaneybound}-\eqnref{eq.vacbound} the critical $\eta$ below which no violation is possible can
be analytically calculated and is given by
\begin{equation}
\label{eq.critampdamp} \tilde{\eta}_{amp} = \frac{8-2^{N+2}+2^N
N(N-1)}{(N-1) \left(2^N N - 8\right)} .
\end{equation}
This is a monotonously decreasing function for $N>2$ and hence the robustness to amplitude damping \textit{decreases} with the number of parties. Considering general detection inefficiencies given by \eqnref{eq.err_spd} and \eqnref{eq.err_sym} for $\hat{\sigma}_{z}$ and $\hat{\sigma}_{x}$ respectively, one obtains \figref{fig.povm_spd_hom}, which also demonstrates that the robustness of the Cabello inequality decreases appreciatively for larger $N$. For the optimal case $N=3$, to obtain a loophole-free test with homodyne realisations of $\hat{\sigma}_{x}$ and SPD realisations of $\hat{\sigma_z}$ we see that $\eta_{spd} > 86.3\%$ is required. For a displacement realisation of $\hat{\sigma}_{x}$ one can show that the bound is $86.4\%$.

\subsection{Tight Bell inequalities}
The inequality (\ref{eq.ineq}) is not a facet of the local
polytope, that is, it is not a tight Bell inequality. So one
should expect that the results obtained in the previous section
are not optimal and that one could improve the thresholds testing
the nonlocality of the W-states by means of tight Bell
inequalities. To probe this we first investigate the tripartite
scenario, which is fully characterized by the Sliwa inequalities
\cite{Sliwa2003}. For higher $N$ one can consider the
WWWZB-inequality~\cite{WWWZB} that fully describes tight Bell
inequalities that can be formed with only full correlators, that
is, only correlations between all subsystems. The WWWZB
inequality describes the multipartite case where each party chooses
between two different dichotomic measurements. There are
$2^{2^{N}}$ tight, linear Bell inequalities which can be succinctly
expressed in terms of the single non-linear inequality
\begin{equation}
\label{WW}
\sum_{r} \left\vert \widehat{\xi}(r)\right\vert \leq1 ,
\end{equation}
where $\widehat{\xi}(r)=2^{-N}\sum_{s} (-1)^{r\cdot s}\left\vert
\xi(s)\right\vert$, $r$ and $s$ are vectors describing a binary
strings, for example $s=(s_1,\cdots,s_N)$ with $s_k=\left\{0,1
\right\}$ and $\xi(s)$ is the full correlator so that $s_k$
indicates the choice of the observable $A_k(s_k)$ at site $k$. In the following we analyze the pure photonic case and the atom-photon case. 

\begin{figure}
\includegraphics[width=.48\textwidth]{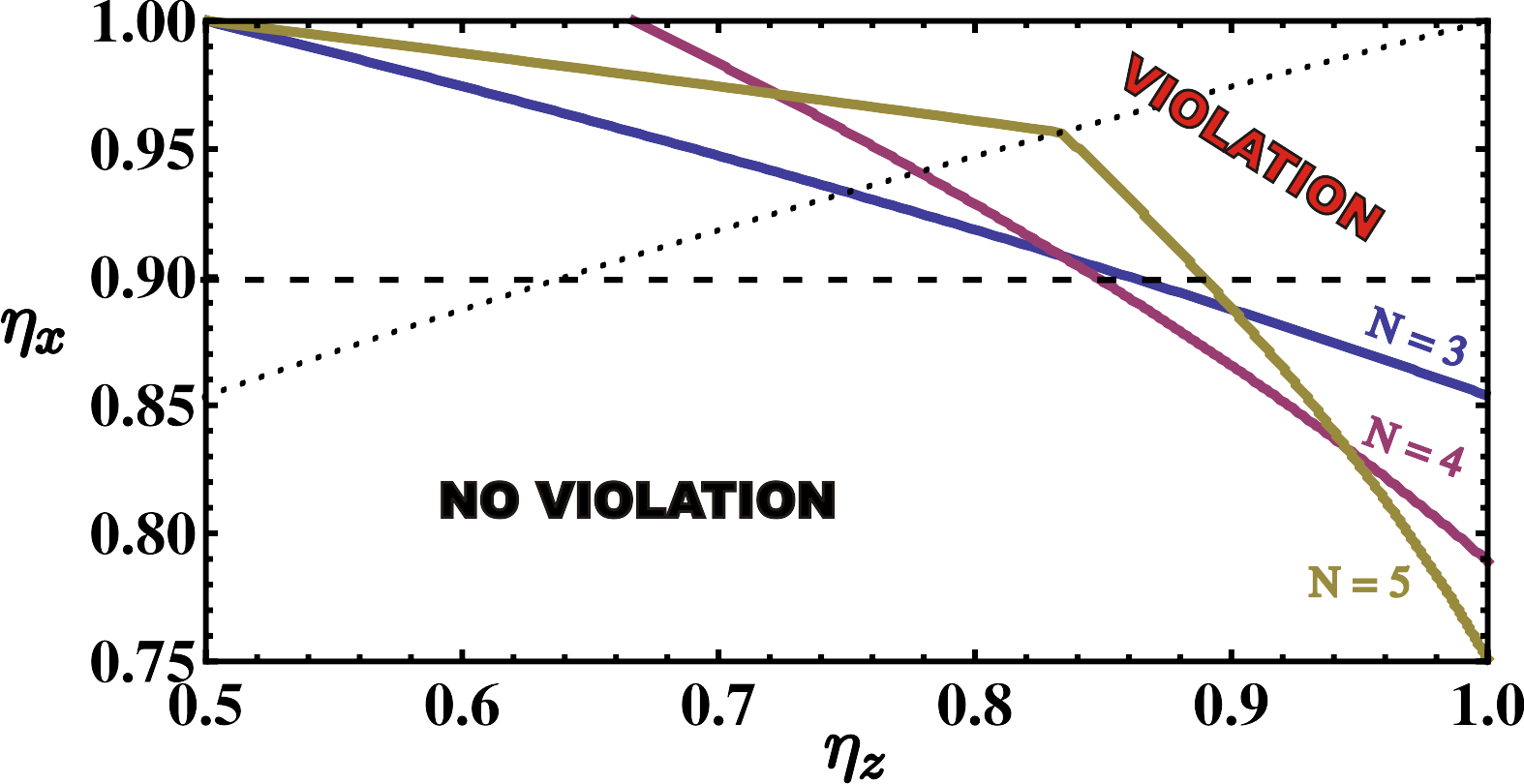}
\caption{Region of violations for the WWWZB inequality using an optical W-state for $N=3,4,5$. The dashed and dotted lines indicate $\eta_x$ as a function of $\eta_z$ for the approximate $\sigma_x$ implementation of \secref{subsec:homo} and amplitude damping respectively. With homodyning, the best critical SPD efficiency is obtained for $N=4\text{ or }7$. When displacements are used, the best is $N=4$.} \label{fig.WWsym}
\end{figure}

\emph{Pure Photonic.} We found that for $N=3$ the Mermin inequality \cite{Mermin1990} - a particular case of the Sliwa and WWWZB inequalities - is the one providing the best robustness to losses and detection inneficiencies. However, for $N \geq 5$ the Mermin inequality is not violated anymore for the chosen measurement settings, even without any imperfections, and instead we consider the criteria (\ref{WW}). The results obtained are displayed in Fig. \ref{fig.WWsym}. It is interesting to note that with WWWZB inequalities, when $\eta_{z}\approx 1$ the required efficiencies for $\eta_{x}$ significantly decrease with increasing $N$. For the particular case of SPD $z$-measurements and approximate $x$-measurements based on ideal homodyning, i.e.~ $\eta_x = \frac{1}{2}(1+\sqrt{2/\pi})$, we find that the best threshold for the SPD efficiency is $\eta_{spd}\thr \approx 85\%$ obtained for $N=4\text{ or }7$. Similarly, when the $x$-measurement is based on displacement followed by SPD, the best threshold is again $85\%$ found for $N=4$. Thus, the performance of the two schemes is similar. However, we note that while $\eta_{hom}=1$ was assumed for the homodyning scheme, there is no such assumption for the displacement scheme. Also, while the former scheme requires a fast switching between homodyning and single-photon detection, in the latter all that is needed is a fast transition between on and off states for the coherent displacement field, which may be easier experimentally (optimal values of $\alpha$ correspond to mean photon numbers on the order of 1). Hence, the displacement scheme may be the more attractive option for implementation.

\begin{figure}
\includegraphics[width=.48\textwidth]{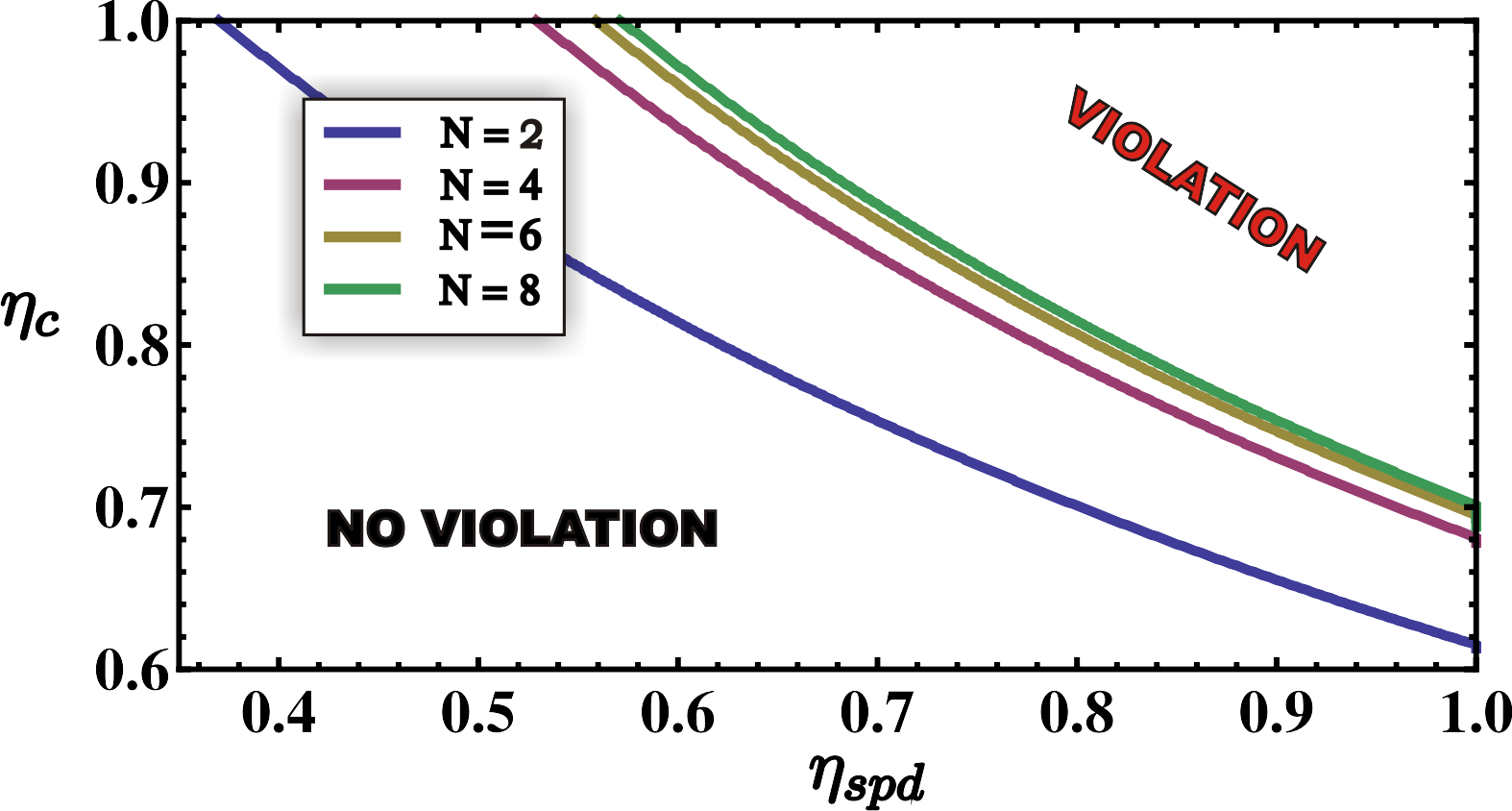}
\caption{Regions of violations of the WWWZB inequality for an optical W-state entangled with an atom for $N=2,4,6,8$ (bottom to top). The approximate $\sigma_x$ implementation of \secref{subsec:homo} with perfect homodyning is assumed. One sees that the optimal case is $N=2$, which implies a very small value for the atom-photon entanglement.}\label{fig.WWasym}
\end{figure}

\emph{Atom-Photon.} Considering the case where the photonic modes are entangled with an atom the thresholds can be considerably improved. We compute the thresholds for violation of (\ref{WW})
in the particular case where $x$-measurements are implemented via ideal homodyning. The results are shown in Fig. \ref{fig.WWasym} for $N=2,4,6,8$. One sees that the optimal thresholds are obtained for $N=2$, a case already considered in Ref.~\cite{Sangouard2011}. For perfect coupling between the atom and the photon, a threshold as low as $\eta_{spd}\thr \approx 37\%$ can be obtained in the limit $\theta \rightarrow 0 $, that is the optimal state is almost separable, in analogy with the Eberhard result~\cite{Eberhard1993}. For $N=3$ on the other hand, the threshold is higher, $\eta_{spd}\thr \approx 55\%$, but the optimal state is more entangled. We find $\theta \approx -0.78 $ which corresponds to a negativity of $\mathcal{N}=0.993$ between the atom and the photonic modes ($\mathcal{N}$ ranges from $0$ for a separable state to $1$ for a maximally entangled state \cite{negativity}). Similar results hold for higher $N$. We note that we are not optimizing over the coefficients that define the photonic state, e.g.~the parameters of beam splitters utilized to divide the single photon between different modes. In principle if such an optimization is done, lower values for the minimum efficiencies can be achieved and in the limit where only one photonic mode is populated we effectively fall back into $N=2$.

\begin{figure}
\includegraphics[width=.48\textwidth]{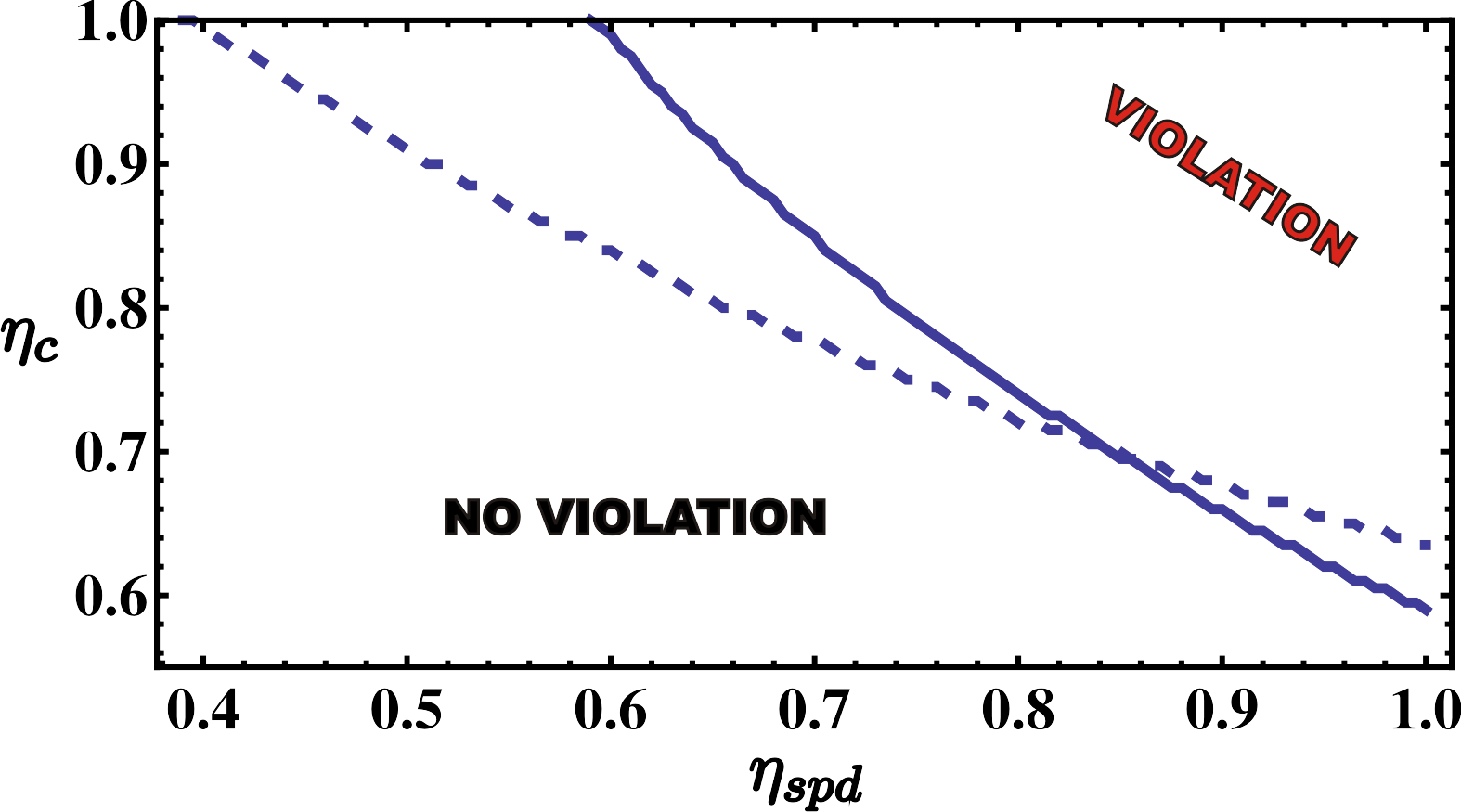}
\caption{Regions of violation of the CHSH inequality. The solid and dashed lines correspond to optical $\sigma_x$-implementations based on displacement and homodyning respectively. For the homodyne detectors $\eta_{hom}=98\%$ while the atoms are detected with efficiency $95\%$.} \label{fig.bipartite}
\end{figure}

To compare the homodyning and displacement schemes for $x$-measurements, we have performed a more detailed calculation for the bipartite case, where the only relevant Bell inequality is CHSH. We first note, that when all detectors as well as the atom-photon coupling are taken to be perfect, the magnitude of the violation is higher in the displacement scheme. The schemes violate the LHV bound of 2 by 2.64 and 2.56. Next, we consider a realistic case in which the detection of the atomic state is imperfect at behaves according to \eqnref{eq.err_spd}, which corresponds e.g.~to a flourescence measurement where the fluorescing state is detected with a finite efficiency. We fix the atomic detection efficiency to be $95\%$. Similarly, we take the homodyne detection efficiency to be $\eta_{hom} = 98\%$. The result is shown in \figref{fig.bipartite}. Under these assumptions, we find that for low SPD efficiency, the best tolerance to coupling loss is obtained using the homodyning scheme, as in Ref.~\cite{Sangouard2011}. On the other hand, for a coupling efficiency below $\sim 70\%$, which may well be the case in experiment, the tolerance to SPD inefficiency is better in the displacement scheme. We remark that the SPD efficiency required for violation for coupling efficiencies around $\sim 65\%$ is high, but that detectors reaching efficiencies in the $90\%$-range have been demonstrated \cite{hadfield2009}.

\section{Polytope}
\label{sec.Polytope}

To probe nonlocality in a more general scenario we consider a
measure of nonlocality based on the Elitzur-Popescu-Rohrlich
(EPR2) decomposition \cite{EPR2}. For a given experiment, that is
for a fixed input state and set of measurement basis, we can
compute the joint probability distribution of the outcomes given
the settings $P = p(o_1,\ldots,o_N|s_1,\ldots,s_N)$. If this
distribution is nonlocal, then the experiment must necessarily
violate some Bell inequality. Conversely, if it is local no Bell
inequality can be violated. Any such $P$ can be decomposed into
the convex mixture of a purely local and a purely nonlocal part
\begin{equation}
P = (1-p_{NL}) P_{L} + p_{NL}P_{NL} ,
\end{equation}
where $P_{L}$ and $P_{NL}$ are respectively a local and a
non-signalling distribution. The minimal value $\tilde{p}_{NL}$
of $p_{NL}$ over all such possible decompositions provides an
unambiguous quantification of the nonlocality, called the {\it
nonlocal content} of $P$.

The nonlocal content can be efficiently calculated by linear
programming \cite{Pitowski1989}, and we make use of this to obtain
critical efficiencies required for detection of nonlocality under
various errors. However, let us first notice that the violation of
any particular Bell inequality allows one to obtain a nontrivial
lower bound on the nonlocal content. Indeed, for any (linear) Bell
inequality $\mathcal{I}\leq\mathcal{I}^L$, the optimal
decomposition of $P$ yields
$\mathcal{I}(P)\equiv(1-\tilde{p}_{NL})
\mathcal{I}(\tilde{P}_{L})+\tilde{p}_{NL}\mathcal{I}(\tilde{P}_{NL})$.
Since $\tilde{P}_{L}$ cannot violate the Bell inequality and
$\mathcal{I}(\tilde{P}_{NL})$ is bounded by its maximal algebraic
value $\mathcal{I}^{max}$, it follows that
\begin{equation}
\tilde{p}_{NL}\geq
\frac{\mathcal{I}(P)-\mathcal{I}^{L}}{\mathcal{I}^{max}-\mathcal{I}^{L}} .
\end{equation}
From here, it follows that any $P$ which saturates the maximal
algebraic value, $\mathcal{I}(P)=\mathcal{I}^{max}$ is
automatically fully nonlocal, i.e. it has $\tilde{p}_{NL}=1$.
This is precisely what happens to GHZ states \cite{GHZ1989}, which
reach the algebraic maximum of the Mermin inequality
\cite{Mermin1990} and are said to be fully nonlocal. We can now
see that the W-state has a similar property for large $N$. On the
one hand, the algebraic maximum of \eqnref{eq.ineq} is
$\mathcal{I}^{max}=1$, since the only positive terms in the
inequality are mutually exclusive events that maximally  sum up to one.
On the other hand, from \eqnref{eq.heaneybound} the W-state violation
approaches 1 exponentially fast, $\mathcal{I}(W)\rightarrow 1$, as
$N\rightarrow\infty$. Thus, the nonlocal content of the W-state
approaches unity.

Let us now turn to investigate the robustness of the nonlocality,
by means of the polytope approach. We will look at the symmetric
W-state \eqnref{eq.W}, for the case of a single-photon
implementation with imperfect photodetection as well as the case
where losses are treated as a third outcome, corresponding e.g.~to
an implementation with polarisation qubits.

\begin{figure}
\includegraphics[width=.48\textwidth]{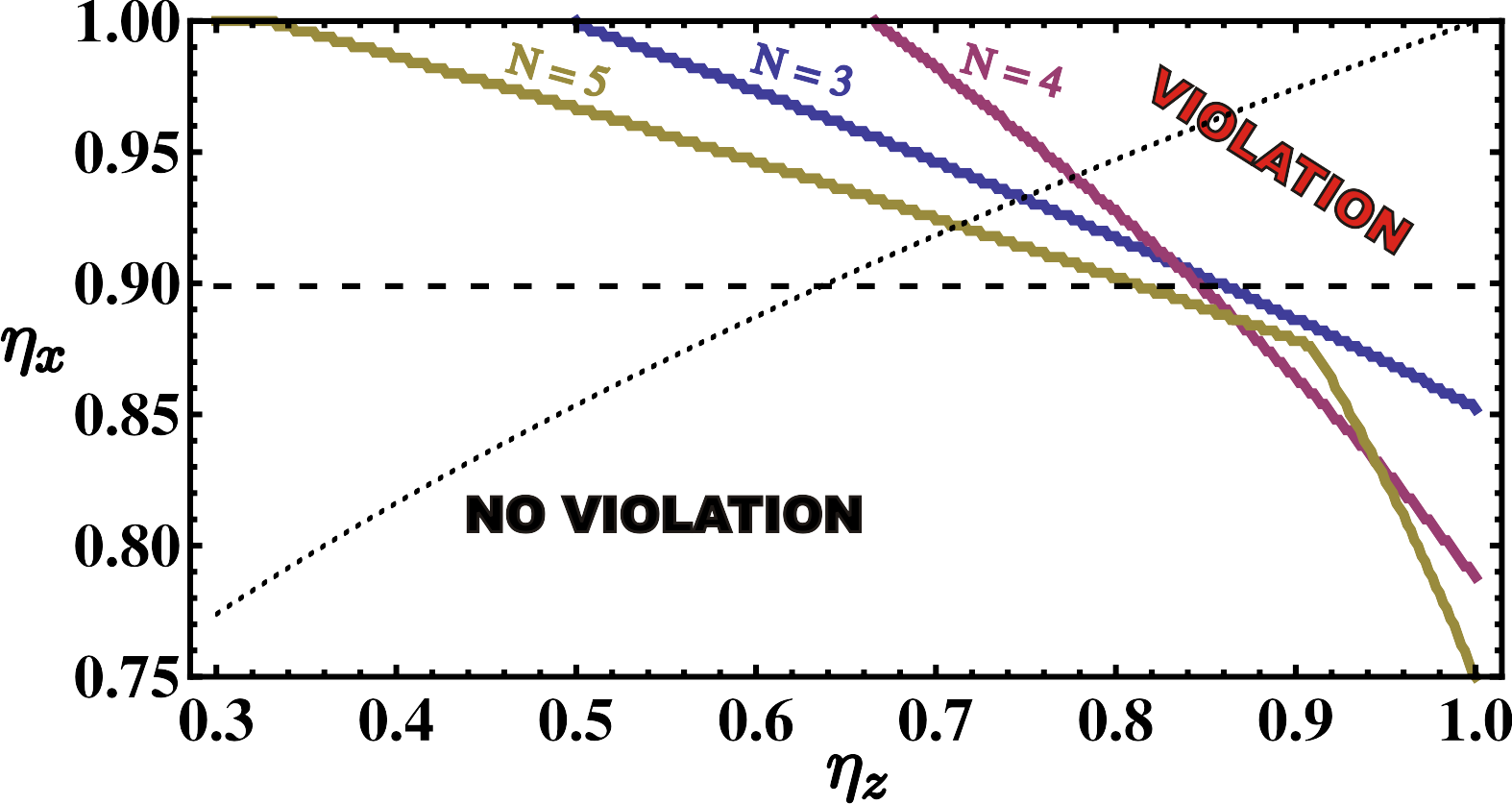}
\caption{Locality regions for the W-state under
$\sigma_z$, $\sigma_x$ measurements with error behaviour given by
\eqnref{eq.err_spd} and \eqnref{eq.err_sym} respectively. Locality
is violated above the solid lines. Intersections of the solid with
the dashed and dotted lines respectively give the critical SPD
efficiencies for the approximate $\sigma_x$ implementation of
\secref{subsec:homo} and for amplitude damping.}
\label{fig.poly_spd_hom}
\end{figure}

\emph{Single-photon case.} We map out the regions of locality when
all parties perform imperfect measurements of $\sigma_z$ and
$\sigma_x$, with the behavior of the imperfections given by
\eqnref{eq.err_spd} and \eqnref{eq.err_sym} respectively. The
result is shown in \figref{fig.poly_spd_hom}, for $N=3,4$, and 5,
and should be compared with \figref{fig.WWsym}. We note that the
regions in the two plots are identical for $N=3,4$, i.e.~the
WWWZB-criterion, based on full correlators only, yields the optimal
solution in these cases. However, for $N=5$, the region obtained
from the full polytope, while qualitatively similar to that found
taking only full correlators, yields lower critical efficiencies
as might be expected in general. One special case of the
\eqnref{eq.err_spd}+\eqnref{eq.err_sym} model is amplitude damping
with ideal detectors, for which $\eta_x = (1+\sqrt{\eta_z})/2$ as
indicated by the dashed line in \figref{fig.poly_spd_hom}. In the
case $N=3$, the threshold for violation agrees with
\eqnref{eq.critampdamp} obtained from the inequality
\eqnref{eq.ineq}, but higher $N$ allows for better tolerance than
this inequality predicts. Another case is the approximate
implementation of $\sigma_x$ given in \secref{subsec:homo}. In
this setting $\eta_x \approx 89.9\%$, which we note falls in a
region with only very modest improvement in critical efficiency up
to $N=5$. It is very interesting to note that within the polytope
approach, when $\eta_x \approx 1$  the required efficiencies for
$\eta_z$ significantly decrease with increasing N and for example
for $N=5$ we have $\eta_z\thr \approx 33\%$.

\emph{Loss as 3rd outcome.} In some cases, while inefficiencies
may not be avoidable it can be possible to detect when a loss
occurs. E.g.~for measurements on optical polarisation qubits, a
no-click event indicates that a photon was lost. For such a
scenario, the no-click event can be treated as an additional
outcome and by considering a polytope for distributions with three
outcomes, the locality regions can be computed. We consider the
ideal W-state \eqnref{eq.W} and two lossy Pauli measurements, the
same for all parties. When the measurement are $\sigma_x$ and
$\sigma_z$, we find that locality is violated whenever $\eta_x >
2(1-\eta_z)$ for $N=3,4$. In particular, in the limit $\eta_z
\rightarrow 1$, any non-zero $\eta_x$ is sufficient for violation.
This is analogous to the result obtained by Garbarino in the
bipartite scenario \cite{garbarino2010}. The same holds when
the two bases are arbitrary, but the same for all parties. Due to
the limitations on our numerical algorithm, we were not able to go
to $N>4$.

\section{Conclusion}

In this paper we have given a detailed analysis of the requirements to perform a nonlocality test using a single-photon W-state shared between multiple parties, and possibly entangled with an additional atomic system. We have focused on implementations using two measurement bases, of which one corresponds exactly to single-photon detection while the other requires and approximate implementation, e.g.~via homodyning. Through a POVM model capturing a broad range of detection imperfections and loss we have numerically obtained the thresholds necessary for a loophole-free violation of local realism. In the case of atom-photon entanglement, comparing different numbers of
parties $N$, we have shown that the bipartite scenario yields the best tolerance to loss. For high atom-photon coupling efficiency, the scheme considered very recently in Ref.~\cite{Sangouard2011}, using both homodyne and single-photon detectors, yields the lowest thresholds for the single-photon detection efficiency, while for lower coupling efficiency another scheme based on displacements and single-photon detection performs better.

With offset in the Bell inequality of Ref.~\cite{Cabello2002} we have demonstrated that for increasing $N$, the local content of the W-state under ideal Pauli measurements tends exponentially fast to zero, that is, the state becomes genuinely nonlocal. Surprisingly, we could also show that despite of that, the robustness of the violation with respect to loss actually decreases rapidly with $N$. Taking a step up in terms of generality, we then considered the WWWZB-inequalities \cite{WWWZB} which provide a compact description of all tight Bell inequalities formed only from full correlators. From these we again found that increasing $N$ does not necessarily lead to better loss thresholds. For example, assuming one approximate Pauli measurement based on ideal homodyning and another based on imperfect single-photon detection, the best threshold is reached already for  $N=4$, giving $\eta_{spd}\thr \approx 85\%$. Finally, we approached the W-state via the polytope of local distributions, covering all possible Bell inequalities and thus taking another step up in terms of generality (as well as numerical complexity). With this approach, we found that the loss thresholds for a single photon are similar to those obtained from WWWZB, though slightly better for $N\leq5$. For implementations that permit the detection of loss events, such as those based on polarisation qubits, much better thresholds can be obtained, with the threshold for one basis approaching 0 when the efficiency in the complementary basis is high.

We note that in the cases where increasing $N$ leads to unchanged or worse loss thresholds, the scaling is not severe. This is a positive result which means that multipartite nonlocality tests become feasible with only slight improvements in detection efficiencies over those required for optimal $N$. As a possible extension of the results presented here, one can consider nonlocality tests involving more than a single photon and that can also be feasible prepared with current technology \cite{Cavalcanti2011}. Also to consider more general scenarios where each party is allowed to perform more than just two measurements \cite{brukner2004} can be a interesting approach to obtain less demanding detection efficiencies.

\textbf{Acknowledgements} We would like to acknowledge helpful discussions with A. Leverrier, as well as with N. Brunner, D. Cavalcanti, and A. Ac\'\i n. R. C. was funded by Q-Essence project. J. B. was funded by the Carlsberg Foundation.


\begin{thebibliography}{99}

\bibitem{Bell1964} J. S. Bell, Physics {\bf 1}, 195 (1964).

\bibitem{Bell_Ions} M.A. Rowe et al., Nature {\bf 409}, 791 (2001); D. N. Matsukevich {\it et al.}, Phys. Rev. Lett. {\bf 100}, 150404 (2008).

\bibitem{Aspect1982} A. Aspect, J. Dalibard, and G. Roger, Phys. Rev. Lett. {\bf 49}, 1804 (1982).

\bibitem{Acin2007} A. Ac\'in, N. Brunner, N. Gisin, S. Massar, S. Pironio, and V. Scarani, Phys. Rev. Lett. {\bf 98}, 230501 (2007).

\bibitem{Pironio2010} S. Pironio {\it et al.}, Nature {\bf 464}, 1021 (2010).

\bibitem{gerhardt2011} I. Gerhardt {\it et al.}, arXiv:1106.3224v2

\bibitem{Pitowski1989} I. Pitowski, {\it Quantum Probability, Quantum Logic} (Springer, Heidelberg, Germany, 1989).

\bibitem{Massar2002} S. Massar {\it et al.}, Phys. Rev. A 66, 052112 (2002).

\bibitem{Cabello2008} A. Cabello, D. Rodr\'iguez, and I. Villanueva, Phys. Rev. Lett. {\bf 101}, 120402 (2008).

\bibitem{Pal2009} K. F. Pal and T. V\'ertesi, Phys. Rev. A {\bf 79}, 022120 (2009).

\bibitem{Vertesi2010} T. V\'ertesi, S. Pironio, and N. Brunner, Phys. Rev. Lett. {\bf 104}, 060401 (2010).

\bibitem{Brunner2007} N. Brunner, N. Gisin, V. Scarani, C. Simon, Phys. Rev. Lett. {\bf 98}, 220403 (2007).

\bibitem{Cabello2007} A. Cabello and J. Larsson, Phys. Rev. Lett. {\bf 98}, 220402 (2007).

\bibitem{Ji2010} S.-W. Ji, J. Kim, H.-W. Lee, M. S. Zubairy, and H. Nha, Phys. Rev. Lett. {\bf 105}, 170404 (2010).

\bibitem{Cavalcanti2011} D. Cavalcanti, N. Brunner, P. Skrzypczyk, A. Salles, and V. Scarani, Phys. Rev. A {\bf 84}, 022105 (2011).

\bibitem{Quintino2011} M. T. Quintino, M. Ara\'ujo, D. Cavalcanti, M. F. Santos, M. T. Cunha, arXiv:1106.2486v1.

\bibitem{Sangouard2011} N. Sangouard {\it et al.}, arXiv:1108.1027v1.

\bibitem{Eberhard1993} P. H. Eberhard, Phys. Rev. A {\bf 47}, R747 (1993).

\bibitem{Patron2004} R. Garc\'ia-Patr\'on, J. Fiur\'a\v{s}ek, N. J. Cerf, J. Wenger, R. Tualle-Brouri, and P. Grangier, Phys. Rev. Lett. {\bf 93}, 130409 (2004).

\bibitem{Wstate} W. D\"ur, G. Vidal, and J. I. Cirac, Phys. Rev. A {\bf 62}, 062314 (2000).

\bibitem{Heaney2011} L. Heaney, A. Cabello, M. F. Santos, V. Vedral, New Journal of Phys. {\bf 13}, 053054 (2011).

\bibitem{Chaves2010} R. Chaves and L. Davidovich, Phys. Rev. A {\bf 82}, 052308 (2010).

\bibitem{CCP} H. Buhrman, R. Cleve, S. Massar, and R. de Wolf, Rev. Mod. Phys. {\bf 82}, 665
(2010).

\bibitem{Laghaout2010} A. Laghaout and G. Bj\"ork Phys. Rev. A {\bf 162}, 033823 (2010).

\bibitem{Cabello2002} A. Cabello, Phys. Rev. A {\bf 65}, 032108 (2002).

\bibitem{Hardy1993} L. Hardy, Phys. Rev. Lett. {\bf 71}, 1665 (1993).

\bibitem{EPR2} A. C. Elitzur, S. Popescu, and D. Rohrlich, Phys. Lett. A {\bf 162}, 25 (1992).

\bibitem{Sliwa2003} C. Sliwa, Phys. Lett. A {\bf 317}, 165 (2003).

\bibitem{Mermin1990} D. N. Mermin, Phys. Rev. Lett. {\bf 65}, 15 (1990).

\bibitem{GHZ1989} D. M. Greenberger, M. A. Horne and A. Zeilinger, in \emph{Bells Theorem, Quantum Theory, and Conceptions of the Universe}, M.
Kafatos (Ed.) (Kluwer Academic Publishers, Dordrecht, The Netherlands, 1989).

\bibitem{ion_review} D. Leibfried, R. Blatt, C. Monroe, and D. Wineland, Rev. Mod. Phys. {\bf 75}, 281 (2003).

\bibitem{optics_review} M. Born  and E. Wolf, \emph{Principles of Optics}, 7th Ed. (Cambridge University Press, 1999).

\bibitem{WWWZB} R. F. Werner and M. W. Wolf, Phys. Rev. A {\bf 64}, 032112 (2001); H. Weinfurter and M. Zukowski, Phys. Rev. A {\bf 64}, 010102(R) (2001); M. Zukowski, C. Brukner, Phys. Rev. Lett. {\bf 88}, 210401 (2002).

\bibitem{negativity} G. Vidal and R. F. Werner, Phys. Rev. A {\bf 65}, 032314 (2002).

\bibitem{hadfield2009} R. H. Hadfield, Nature Photonics {\bf 3}, 696 (2009).

\bibitem{garbarino2010} G. Garbarino, Phys. Rev. A {\bf 81}, 032106 (2010).

\bibitem{brukner2004} W. Laskowski, T. Paterek, M. Zukowski, and C. Brukner, Phys. Rev. Lett. \textbf{93}, 200401 (2004).

\end{thebibliography}
\end{document}